# Delta-like singularity in the Reduction of the Laplace Equation for Spherical Coordinates and the Validity of Radial Schrodinger Equation

Anzor A.Khelashvili[1,2] and Teimuraz P. Nadareishvili[1]

[1] *Inst. of High Energy Physics, Iv. Javakhishvili Tbilisi State University, University Str. 9, 0109, Tbilisi, Georgia*

[2] *St.Andrea the First-called Georgian University of Patriarchy of Georgia, Chavchavadze Ave.53a, 0162, Tbilisi, Georgia.*

Corresponding author. Phone: 011+995-98-54-47 ; E-mail address: teimuraz.nadareishvili@tsu.ge;

**Abstract:** By careful exploration of separation of variables into the Laplacian in spherical coordinates, we obtain the extra delta-like singularity, elimination of which restricts the radial wave function at the origin. This constraint has the form of boundary condition for the radial Schrodinger equation.

## 1. Introduction.

It is well known that the Laplace operator appears in many places of physical as well as of mathematical problems. Especially in quantum mechanics the dynamics of any physical system is described by the three dimensional Schrodinger equation [1,2]

$$\Delta \psi(\vec{r}) + 2m[E - V(r)]\psi(\vec{r}) = 0 \qquad (1)$$

In the most interesting physical problems the central potential $V(\vec{r}) = V(|\vec{r}|) \equiv V(r)$ is frequently encountered, therefore reduction to the one-dimensional (radial) equation is the wide-spread procedure.

The traditional way is the application of the substitution $\psi(\vec{r}) = R(r) Y_l^m(\theta, \varphi)$, where $Y_l^m(\theta, \varphi)$ is the spherical harmonics and because of the continuity and uniqueness, orbital quantum numbers $l$ are integers, $l = 0,1,2,...$, whereas $m = -|l|,...,l$. After this substitution angular variables are separated and we are left to the equation for the full radial function $R(r)$

$$\frac{d^2 R}{dr^2} + \frac{2}{r}\frac{dR}{dr} + 2m[E - V(r)]R - \frac{l(l+1)}{r^2}R = 0 \qquad (2)$$

It is traditional trick in quantum mechanics to avoid the first derivative term from this equation by substitution

$$R(r) = \frac{u(r)}{r} \qquad (3)$$

after which a naïve calculation gives the equation for the new radial wave function $u(r)$ in the form

$$\frac{d^2 u(r)}{dr^2} - \frac{l(l+1)}{r^2} u(r) + 2m[E - V(r)]u(r) = 0 \qquad (4)$$

Just this equation plays an important role in quantum mechanics since its birth. However, as is clarified in recent years, there is an ambiguity in derivation of boundary condition for $u(r)$ at the origin $r = 0$, especially in case of singular potentials [3-5].

According to this reason many authors content themselves by consideration only a square integrability of radial function and do not pay attention to its behavior at the origin. Of course, this is permissible mathematically and the strong theory of linear differential operators allows for such approach [6-8]. There appears so-called Self-Adjoint Extended (SAE) physics [9], in the framework of which among physically reasonable solutions one encounters also many curious results, such as bound states in case of repulsive potential [10] and so on. We think that these highly unphysical results are caused by the fact that without suitable boundary condition at the origin a functional domain for radial Schrodinger Hamiltonian is not restricted correctly [11].

Careful investigation, performed below, shows that the validity of radial equation (4) is not correctly established. Indeed, it is physically (and mathematically, of course) warranted that the equation obtained after separation of variables, must be compatible with the primary equation. It is necessary condition for the correctness of a separation procedure.

## 2. Rigorous derivation of radial equation.

In case of reduction of Laplace operator the transition from Cartesian to spherical coordinates is not unambiguous, because the Jacobean of this transformation [12] $J = r^2 \sin\theta$ is singular at $r = 0$ and $\theta = n\pi (n = 0,1,2,...)$. Angular part is fixed by the requirement of continuity and uniqueness. This gives the unique spherical harmonics $Y_l^m(\theta, \varphi)$ mentioned above.

Note that in the reduction of Laplace operator usually is pointed out that $r > 0$. However $\vec{r} = 0$ is an ordinary point in full Schrodinger equation (1), but it is a point of singularity in the reduction of variables. Thus, the knowledge of specific boundary behavior is necessary. We underline that the equation (2) is correct, but the substitution (3) enhances singularity at $r = 0$ and may cause some misunderstandings.

Indeed, let us rewrite the full radial equation (2) after this substitution

$$\frac{1}{r}\left(\frac{d^2}{dr^2} + \frac{2}{r}\frac{d}{dr}\right)u(r) + u(r)\left(\frac{d^2}{dr^2} + \frac{2}{r}\frac{d}{dr}\right)\left(\frac{1}{r}\right) + \\ + 2\frac{du}{dr}\frac{d}{dr}\left(\frac{1}{r}\right) - \left[\frac{l(l+1)}{r^2} - 2m(E - V(r))\right]\frac{u}{r} = 0 \qquad (5)$$

We write equation in this form deliberately, indicating action of radial part of Laplacian on relevant factors explicitly. It seems that the first derivatives of $u(r)$ cancelled and we are faced to the following equation

$$\frac{1}{r}\left(\frac{d^2 u}{dr^2}\right) + u\left(\frac{d^2}{dr^2} + \frac{2}{r}\frac{d}{dr}\right)\left(\frac{1}{r}\right) - \frac{l(l+1)}{r^2}\frac{u}{r} + 2m(E - V(r))\frac{u}{r} = 0 \quad (6)$$

Now if we differentiate the second term "naively", we'll derive zero. But it is true only in case, when $r \neq 0$. However, below we show that in general this term is proportional to the 3-dimensional delta function. Indeed, taking into account that,

$$\frac{d^2}{dr^2} + \frac{2}{r}\frac{d}{dr} = \frac{1}{r^2}\frac{d}{dr}\left(r^2 \frac{d}{dr}\right) \equiv \Delta_r \quad (7)$$

is the radial part of the Laplace operator and therefore [13]

$$\Delta_r\left(\frac{1}{r}\right) = \Delta\left(\frac{1}{r}\right) = -4\pi\delta^{(3)}(\vec{r}) \quad (8)$$

we obtain the equation for $u(r)$

$$\frac{1}{r}\left[-\frac{d^2 u(r)}{dr^2} + \frac{l(l+1)}{r^2} u(r)\right] + 4\pi\delta^{(3)}(\vec{r})u(r) - 2m[E - V(r)]\frac{u(r)}{r} = 0 \quad (9)$$

We see that there appears the extra delta-function term. It's presence in the radial equation is physically nonsense and must be eliminated. Note that when $r \neq 0$, this extra term vanishes owing to the property of the delta function and if, in this case, we multiply this equation on $r$, we'll obtain the ordinary radial equation (4).

However if $r = 0$, multiplication on $r$ is not permissible and this extra term remains in Eq. (9). Therefore one has to investigate this term separately and find another ways to abandon it.

The term with 3-dimensional delta-function must be comprehended as being integrated over $d^3 r = r^2 dr \sin\theta d\theta d\varphi$. On the other hand [13]

$$\delta^{(3)}(\vec{r}) = \frac{1}{|J|}\delta(r)\delta(\theta)\delta(\varphi) \quad (10)$$

Taking into account all the above mentioned relations, one is convinced that extra term still survives, but now in the one-dimensional form

$$u(r)\delta^{(3)}(\vec{r}) \rightarrow u(r)\delta(r) \quad (11)$$

Its appearance as a point-like source, breaks many fundamental principles of physics, which is not desirable. The only reasonable way to remove this term without modifying Laplace operator or including compensating delta function term in the potential $V(r)$, is to impose the requirement

$$u(0) = 0 \quad (12)$$

(note, that multiplication of Eq. (9) on $r$ and then elimination this extra term owing the property $r\delta(r) = 0$ is not legitimated procedure, because effectively it is equivalent to multiplication on zero).

Therefore we conclude that the radial equation (4) for $u(r)$ is compatible with the full Schrodinger equation (1) if and only if the condition $u(0) = 0$ is fulfilled. *The radial equation (4) supplemented by the condition (12) is equivalent to the full Schrodinger equation (1).* It is in accordance with the Dirac requirement [2], that the solutions of the radial equation must be compatible with the full Schrodinger equation. It is remarkable to see that the supplementary condition (12) has a form of boundary condition at the origin.

## 3. Comments, some applications and conclusions

Some comments are in order here: equation for $R(r) = \dfrac{u(r)}{r}$ has its usual form (2). Derivation of boundary behavior from this equation is as problematic as for $u(r)$ from Eq. (4). Problem with delta function arises only in the course of elimination of the first derivative. Now, after the condition (12) is established, it follows that the full wave function $R(r)$ is less singular at the origin than $r^{-1}$. Though, this conclusion could be hasty because the transition to Eq. (4) for $R(r)$ is not necessary. It is also remarkable to note that the boundary condition (12) is valid whether potential is regular or singular. It is only consequence of particular transformation of Laplacian. Different potentials can only determine the specific way of $u(r)$ tending to zero at the origin and the delta function arises in the reduction of the Laplace operator every time. All of these statements can easily be verified also by explicit integration of Eq. (9) over a small sphere with radius $a$ tending it to zero at the end of calculations.

It seems very curious that this fact was unnoticed up by physicists till now in spite of numerous discussions [14].

Apparently mathematicians knew about singular behavior of Laplace operator for a long time. But their results did not find a relevant representation in physical literature, while the delta function became popular after Dirac. Therefore the fact, described above, seems to us as being very curious.

We discuss another important point with regard to radial Laplacian. It is well known from some books on special functions that there is the following operator relation [13]

$$\Delta_r = \frac{1}{r^2}\frac{d}{dr}\left(r^2 \frac{d}{dr}\cdot\right) = \frac{1}{r}\frac{d^2}{dr^2}(r\cdot) \qquad (13)$$

Here the dot denotes the action of this expression on some function. The validity of this relation is easily verified by direct calculation. But this equality fails at point $r = 0$. Indeed, let us act by both sides on the full radial function $R(r)$:

$$\frac{1}{r^2}\frac{d}{dr}\left(r^2 \frac{dR}{dr}\right) = \frac{1}{r}\frac{d^2}{dr^2}(rR) = \frac{1}{r}\frac{d^2}{dr^2}u(r) \qquad (14)$$

Exactly this relation is used in mathematical literature for special functions [15].

If it will be true everywhere then there does not appear any problem in derivation of the radial equation. But now we know that after substitution of $R(r) = \frac{u(r)}{r}$ on the left-hand side it follows

$$\frac{1}{r^2}\frac{d}{dr}\left(r^2 \frac{d}{dr}\frac{u}{r}\right) = \frac{1}{r}\frac{d^2 u}{dr^2} - 4\pi\delta(\vec{r})u \qquad (15)$$

Therefore previous operator equality must be modified perhaps as follows

$$\frac{1}{r^2}\frac{d}{dr}\left(r^2 \frac{d}{dr}\cdot\right) = \frac{1}{r}\frac{d^2}{dr^2}(r\cdot) - 4\pi\delta^{(3)}(\vec{r})r\cdot \qquad (16)$$

This relation is correct at every point including the origin. Validity of this relation may be checked by acting on $R(r)$.

The relation $u(0) = 0$ is not only the boundary condition for the radial equation, but it is relation which must be necessarily fulfilled in order to have the radial equation in its usual form compatible to the full Schrodinger equation. Accidentally it has a boundary condition form. Without this condition the radial equation is not valid.

Now, that this condition has been established, many problems can be considered rigorously by taking it into account. Remarkably, all the results obtained earlier for regular potentials with the boundary condition (12) remain unchanged. In the most textbooks on quantum mechanics $r \to 0$ behavior is obtained from Eq. (4) in case of regular potentials. When equation like (4) is known, the derivation of boundary behavior from it is almost trivial procedure. It depends on the behavior of potential under consideration.

But we have shown that this equation takes place only together with boundary condition (12). On the other hand, for *singular potentials* this condition will have far-reaching implications. Many authors neglected boundary condition entirely and were satisfied only by square integrability. But this treatment, after leakage into the forbidden regions and through a self-adjoint extension procedure, sometimes yields curious unphysical results. Below we consider some simple consequences, showing the differences, which arise with and without above mentioned condition:

(i) Regular potentials at the origin:

$$\lim_{r \to 0} r^2 V(r) = 0 \qquad (17)$$

In this case, after substitution at the origin $u \sim r^a$, it follows from indicial equation, that $a(a-1) = l(l+1)$, which gives two solutions $u \underset{r \to 0}{\sim} c_1 r^{l+1} + c_2 r^{-l}$ (see, any textbooks on quantum mechanics). For non-zero $l$-s the second solution is not square integrable and is ignored usually. But for $l = 0$, many authors discuss how to deal with this solution [16], which is also square integrable at origin. According to condition (12), this solution must be ignored.

(ii) Transitive attractive singular potentials at the origin:

$$\lim_{r \to 0} r^2 V(r) = -V_0 = const; \quad V_0 > 0 \qquad (18)$$

In this case, the indicial equation takes form $a(a-1) = l(l+1) - 2mV_0$, which has two solutions: $a = \frac{1}{2} \pm \sqrt{\left(l + \frac{1}{2}\right)^2 - 2mV_0}$. Therefore

$$\underset{r \to 0}{u} \sim c_1 r^{\frac{1}{2}+P} + c_2 r^{\frac{1}{2}-P} \quad ; \quad P = \sqrt{\left(l + \frac{1}{2}\right)^2 - 2mV_0} \tag{19}$$

It seems, that both solutions are square integrable at origin as long as $0 \le P < 1$. Exactly this range is studied in most papers (see for example [10]), whereas according to boundary condition (12) we have $0 \le P < \frac{1}{2}$. The difference is essential. Indeed, the radial equation has form

$$u'' - \frac{P^2 - 1/4}{r^2} u = 2mE \tag{20}$$

Depending on whether $P$ exceeds $1/2$ or not, the sing in front of the fraction changes and one can derive attraction in case of repulsive potential and vice versa. Boundary condition (12) avoids this unphysical region $\frac{1}{2} \le P < 1$.

Lastly, we note that the same holds for radial reduction of the Klein-Gordon equation, because in three dimensions it has the following form

$$\left(-\Delta + m^2\right)\psi(\vec{r}) = [E - V(r)]^2 \psi(\vec{r}) \tag{21}$$

and the reduction of variables in spherical coordinates will proceed to absolutely same direction as in Schrodinger equation.

*Acknowledgements.* We want to thank Profs. John Chkareuli, Sasha Kvinikhidze and Parmen Margvelashvili for valuable discussions. One of us (A.Kh.) is indebted to thank Prof. Boris Arbuzov for reading the manuscript.